\documentclass[aps,preprint]{revtex4}
\usepackage{graphicx}
\usepackage{epsfig}
\usepackage{amsmath, amssymb}

\begin{document}

\title{ Kerr--Schild--AdS  geometries in quadratic $f(R)$ gravity: \mbox{A no-go theorem} }
\author{\large Alikram N. Aliev}
\address{Department of Basic Sciences, Faculty of Engineering and Natural Sciences,
Maltepe University, 34857 Maltepe, Istanbul, T\"{u}rkiye}

\date{\today}

\begin{abstract}

We investigate Kerr--Schild--AdS geometries in quadratic \(f(R)\) gravity without imposing the constant-curvature condition \(R=R_0\) \emph{a priori}. We show that the field equations dynamically enforce constant scalar curvature and uniquely select the Kerr--AdS family of solutions. Thus, quadratic \(f(R)\) gravity admits no Kerr--AdS solutions beyond the Einstein branch, establishing a no-go theorem for this class of geometries.

\end{abstract}

\pacs{04.20.Cv, 04.50.+h}
\maketitle
\newpage

\section{Introduction}

The Kerr--Schild ansatz for the spacetime metric has played a central role in constructing the general stationary and axisymmetric Kerr--Newman family of solutions that describe rotating charged black holes in general relativity \cite{ks}. The defining property of the Kerr--Schild approach is that the exact metric can be written in a form closely resembling its linearized counterpart around a flat background spacetime. This structure leads to a remarkable simplification of the field equations, thereby reducing them to a set of effectively linear equations. As a result, the Kerr–Schild framework provides a powerful tool for constructing exact solutions while fully preserving the nonlinear structure of the underlying theory.  In Ref.~\cite{glpp}, it was shown that the four-dimensional Kerr--de Sitter solution, originally discovered by Carter \cite{carter}, can be expressed in Kerr--Schild form on a de Sitter (or anti--de Sitter) background spacetime. Within this framework, the authors further showed that the Kerr--Schild--de Sitter geometries are exact solutions of the Einstein equations in arbitrary dimensions. Subsequent developments extended the Kerr--Schild--de Sitter framework beyond general relativity, including applications to both gauged and ungauged supergravity theories \cite{ad, bkastor, sqing, kubiznak}.

These developments naturally raise  the question of whether, and to what extent, the remarkable properties of the Kerr--Schild framework persist in modified theories of gravity. 
In recent years, such theories have attracted significant attention as possible extensions of general relativity, motivated by both quantum corrections and cosmological applications. 
Among them, $f(R)$ gravity is of particular interest, since it provides one of the simplest and most consistent extensions of general relativity, in which the Einstein--Hilbert action is replaced by a nonlinear function of the Ricci scalar $R$ \cite{star, soti, nojiri}. Black hole solutions in this theory are of considerable importance for understanding both its mathematical structure and physical implications. However, the construction of such solutions is highly nontrivial due to the nonlinear nature of the field equations. A particularly important role is played by the constant-curvature condition $R=R_0$, under which the field equations reduce to the Einstein equations with an effective cosmological constant. In this case, after an appropriate redefinition of the metric parameters, the familiar Kerr--AdS geometries of general relativity are recovered as solutions of $f(R)$ gravity \cite{larran, cembra}. 

This observation leads to a more specific question: What happens to the Kerr--AdS geometries in \(f(R)\) gravity if the constant-curvature condition \(R=R_0\) is not imposed \emph{a priori}?

The purpose of this Letter is to answer this question by examining the corresponding Kerr--Schild framework in the context of quadratic \(f(R)\) gravity. We show that all deviations from constant curvature are controlled by a single radial function. The field equations then dynamically enforce constant scalar curvature and uniquely select the Kerr--AdS family of solutions, thereby establishing a no-go theorem beyond the Einstein branch.

\section{Field equations and the Kerr--Schild ansatz}

The field equations of $f(R)$ gravity follow from the variation of the action \cite{star, soti, nojiri}
\begin{eqnarray}
S &=&  \frac{1}{16 \pi G} \int d^4 x \sqrt{-g}\,f(R) 
 \label{action}
\end{eqnarray}
with respect to the metric tensor $g_{\mu\nu}$. The resulting equations take the form
\begin{eqnarray}
 &&  f'(R) R_{\mu\nu}  - \frac{1}{2}\, f(R)\,  g_{\mu\nu} + \left( g_{\mu\nu} \, 
 \nabla^2 - \nabla_\mu \nabla_{\nu} \right) f'(R)    =  0,
 \label{eq1}
\end{eqnarray}
where $f'(R)= df/dR$ and  the operator $\nabla_{\mu}$ denotes covariant differentiation. 
Taking the trace of Eq.~(\ref{eq1}) yields
\begin{eqnarray}
3\, \nabla^{2} f'(R) + R f'(R)  &=& 2  f(R).
\label{treq1}
\end{eqnarray}

We now specialize to the quadratic $f(R)$ model, given by
\begin{eqnarray}
f(R) = R + \alpha R^2 - 2\Lambda,
 \label{quadratic}
\end{eqnarray}
where \(\alpha\) is the quadratic coupling constant  and \(\Lambda\) is the cosmological constant.  In this case, the field equations take the form
\begin{eqnarray}
E_{\mu\nu}& \equiv &f'(R) R_{\mu\nu} - \frac{1}{2}\,  g_{\mu\nu} \left( R+\alpha R^2  - 2 \Lambda \right)  + \left( g_{\mu\nu} \, 
 \nabla^2 - \nabla_\mu \nabla_{\nu} \right) f'(R)   = 0
 \label{eq2}
\end{eqnarray}
where  $f'(R)= 1 + 2 \alpha R $. The corresponding trace equation becomes
\begin{eqnarray}
6\alpha  \nabla^2 R - R + 4\Lambda = 0.
\label{treq2}
\end{eqnarray}

Next, we consider the Kerr--Schild metric form \cite{glpp}
\begin{equation}
ds^{2}=d\bar{s}^{2}+ H\, k\otimes k\,,
\label{gks1}
\end{equation}
where $ d\bar{s}^{2} $ is the background AdS spacetime metric, which in oblate spheroidal coordinates takes the form
\begin{equation}
d\bar{s}^2 =
-\left(1+\frac{r^2}{l^2}\right) \frac{\Delta_\theta}{\Xi}\,dt^2
+\left(1+\frac{r^2}{l^2}\right)^{-1}  \frac{\Sigma}{r^2+a^2} \,dr^2
+\frac{\Sigma}{\Delta_\theta}\,d\theta^2
+\frac{(r^2+a^2)\sin^2\theta}{\Xi}\,d\phi^2\,.
\label{adsb}
\end{equation}
Here
\begin{eqnarray}
\Delta_\theta & = & 1 -\frac{a^2}{l^2} \,\cos^2\theta \,,~~~~~
\Xi =1 - \frac{a^2}{l^2}\,,~~~~~
\Sigma  =  r^2+ a^2 \cos^2\theta \,,
\label{ksmetfunc}
\end{eqnarray}
$a$ is  the oblate spheroidal parameter, \(H(r,\theta)\) is a scalar (profile) function and \(k_\mu\) is a null vector with respect to both the background and the full spacetime metrics. 
For convenience, we also introduce the curvature radius $l$, related to the negative AdS background curvature  through
\begin{eqnarray}
l^{-2} = -\frac{R_0}{12}\,.
\label{ll}
\end{eqnarray}
The Kerr--Schild one-form is chosen as
\begin{eqnarray}
k_\mu dx^\mu = \frac{\Delta_\theta}{\Xi}\,dt
+ \left(1+\frac{r^2}{l^2}\right)^{-1}
\frac{\Sigma}{r^2+a^2} \,dr
- \frac{a\sin^2\theta}{\Xi}\,d\phi,
\label{ks1form}
\end{eqnarray}
so that
\[
k_\mu k^\mu = 0,
\qquad
k^\nu \nabla_\nu k^\mu = 0,
\]
showing that $k^\mu$ is both null and geodesic. 

Motivated by the Kerr--Schild--AdS geometry, we consider the ansatz
\begin{equation}
H(r,\theta)=\frac{m(r)}{\Sigma},
\label{subclass}
\end{equation}
which preserves the characteristic oblate spheroidal structure of the spacetime while allowing for a nontrivial radial deformation of the Kerr--AdS profile function. This choice retains the essential geometric features of the underlying Kerr--AdS geometry and renders the field equations analytically tractable.

It is straightforward to show that for this ansatz the scalar curvature takes the remarkably simple form
\begin{equation}
R-R_0=\frac{m''(r)}{\Sigma}\,,
\label{riccisc}
\end{equation}
where \(R_0\) refers to the background AdS spacetime as defined in Eq.~(\ref{ll}) and the prime denotes differentiation with respect to \(r\). At this stage, the system does not yet reduce to the Einstein branch. Instead, the scalar curvature remains dynamical  and the trace equation alone in Eq.~(\ref{treq2}) is insufficient to determine the solution. As we shall see, the decisive restrictions arise from the off-diagonal components of the field equations. Their consistency conditions dynamically force the system onto the constant-curvature branch, thereby selecting the Kerr--AdS family of solutions.

\section{No-go theorem}

Our main result is summarized in the following no-go theorem.

\medskip
\noindent
\textbf{Theorem.}
\emph{Consider Kerr--Schild--AdS geometries of the form (\ref{subclass})  subject to the vacuum field equations of quadratic \(f(R)\) gravity in Eq.~(\ref{eq2}), without imposing the constant-curvature condition \mbox{$R=R_0$} \emph{a priori}. Then the field equations dynamically enforce constant scalar curvature and uniquely select the Kerr--AdS family of solutions. Consequently, the theory admits no rotating solutions beyond the Einstein branch.}

The proof of the theorem follows from the off-diagonal components of the field equations in Eq.~(\ref{eq2}). Among them, the  \(r\theta\) component plays a decisive role and provides the first nontrivial consistency condition. A direct computation gives
\begin{equation}
E_{r\theta} = - \frac{3 a^2 \alpha \sin2\theta \left[ -8 r\,m''(r)+2 \Sigma\,m'''(r)\right]}{2 \Sigma^3}\,,
\label{erth1}
\end{equation}
where \(\Sigma\) is defined in Eq.~(\ref{ksmetfunc}).  For \(a\neq0\) and away from the symmetry axes, \(\sin 2\theta \neq0\), the field equation \(E_{r\theta}=0\) implies
\begin{equation}
-8r\,m''(r)+2\Sigma\,m'''(r)=0.
\label{conds1}
\end{equation}
Since \(m(r)\) depends only on the radial coordinate, whereas \(\Sigma\) contains explicit angular dependence, Eq.~(\ref{conds1}) can hold for all values of \(\theta\) only if
\begin{equation}
m'''(r)=0.
\label{m3}
\end{equation}
Substituting this condition back into Eq.~(\ref{conds1}) then gives
\begin{equation}
m''(r)=0.
\label{m2}
\end{equation}
Consequently, substituting Eqs.~(\ref{m3}) and (\ref{m2}) into the trace equation (\ref{treq2}) yields
\begin{equation}
m^{(4)}(r)=0 .
\label{m4}
\end{equation}
This result is fully consistent with the conditions obtained above. We therefore arrive at 
the set of consistency conditions
\begin{equation}
m''(r)=m'''(r)=m^{(4)}(r)=0.
\label{condsg}
\end{equation}
After imposing the conditions in Eq.~(\ref{condsg}), the remaining off-diagonal components reduce to
\begin{eqnarray}
E_{tr} &=& -\frac{(l^2-24 \alpha) \Delta_\theta\,m(r)}
{\Xi  (r^2+a^2)(r^2+l^2)  \Sigma^2}  \, \mathcal{K}(r) , ~~~~ E_{r \phi}= - E_{tr}\, \frac{a \sin^2 \theta}{\Delta_\theta} , \nonumber\\ [3mm]
E_{t \phi} &=& - \frac{a \sin^2 \theta (l^2-24 \alpha)  \Delta_\theta \,  \left[2(r^2+a^2)(r^2+l^2)-l^2 m(r)\right]}{l^4\Xi^2\Sigma^3}\,  \mathcal{K}(r)\,,
\label{offdiageqs}
\end{eqnarray}
where we have introduced the common factor
\begin{equation}
\mathcal{K}(r)= r m'(r) - m(r)\,.
\label{common}
\end{equation}
Furthermore,  a direct computation shows that  the diagonal components $ E_{tt}, E_{rr}, E_{\theta\theta}$  and $E_{\phi\phi} $  are all proportional to the same radial factor \(\mathcal K(r)\). Therefore, the remaining field equations consistently reduce to
\begin{equation}
r m'(r) - m(r)=0,
\label{rmprime}
\end{equation}
whose unique solution is
\begin{equation}
m(r)=Cr,
\label{mlinear}
\end{equation}
where \(C\) is an integration constant. Substituting Eq.~(\ref{mlinear}) into the Kerr--Schild--AdS ansatz reproduces the Kerr--AdS family of solutions. Hence, the field equations dynamically enforce constant scalar curvature and uniquely select the Kerr--AdS family, leaving no rotating solutions beyond the Einstein branch. This completes the proof.

\section{Discussion}

The main result of this work is that the constant-curvature condition is not imposed but rather emerges dynamically from the field equations themselves. The decisive consistency condition originates from the off-diagonal \(r\theta\) component of the field equations of quadratic \(f(R)\) gravity. Remarkably, this component is proportional to the quadratic coupling constant \(\alpha\) and therefore vanishes identically in the Einstein limit \(\alpha=0\). Thus, the condition in Eq.~(\ref{m2}) is not a constraint inherited from general relativity, but rather a consistency requirement generated by the \(R^2\) term of the theory. The resulting consistency conditions force the  profile function to satisfy Eqs.~(\ref{m3}) and (\ref{m2}), which in turn imply \(R=R_0\).  Therefore,  the Einstein branch is dynamically selected by the higher-curvature corrections.

A second noteworthy feature is that, once the consistency conditions obtained from the \(r\theta\) equation are imposed, all remaining field equations collapse onto the single radial constraint given by Eq.~(\ref{rmprime}). Consequently, the profile function is uniquely fixed to be linear, as in Eq.~(\ref{mlinear}), and the Kerr--AdS family of solutions is recovered. In this sense, the quadratic curvature correction does not generate a new rotating branch within the Kerr--Schild framework; instead, it drives the system back to the Einstein branch.

It is important to emphasize the scope of the result.  The ansatz (\ref{subclass}) is not an arbitrary truncation, but is directly motivated by the Kerr--AdS geometry, preserving the characteristic oblate spheroidal structure while allowing the radial profile to be determined dynamically by the field equations. Extensions including electric charge appear to be straightforward, while it would be interesting to investigate whether analogous obstructions persist in other higher-curvature theories.

Finally, although our analysis has been performed for the quadratic \(f(R)\) model, the mechanism underlying the no-go theorem appears to be more general. For the Kerr--Schild ansatz (\ref{subclass}), all deviations from constant curvature are controlled by the radial quantity \(m''(r)\). Since the higher-derivative contributions in a generic \(f(R)\) theory involve derivatives of the scalar curvature, the structure of the field equations suggests that the off-diagonal consistency conditions may continue to dynamically enforce \(m''(r)=0\), thereby driving the system onto the constant-curvature branch.

\end{document}